\documentclass{elsart}

\usepackage{epsfig}
\begin{document}
\def\beq{\begin{equation}}
\def\eeq{\end{equation}}
\def\bea{\begin{eqnarray}} 
\def\eea{\end{eqnarray}}
\def\eps{\epsilon}
\newcommand{\ket}[1]{| #1 \rangle}
\newcommand{\bra}[1]{\langle #1 |}
\newcommand{\braket}[2]{\langle #1 | #2 \rangle}
\newcommand{\proj}[1]{| #1\rangle\!\langle #1 |}
\newcommand{\ba}{\begin{array}}
\newcommand{\ea}{\end{array}}

\begin{frontmatter}


\title{Detecting Quantum Entanglement}


\author{Barbara M. Terhal}

\address{IBM Watson Research Center,
P.O. Box 218, Yorktown Heights, NY 10598, USA\thanksref{email}}
\thanks[email]{E-mail:terhal@watson.ibm.com}

\begin{abstract}
We review the criteria for separability and quantum entanglement, both in 
a bipartite as well as a multipartite setting. We discuss Bell inequalities, entanglement witnesses, entropic inequalities, bound entanglement and several features of multipartite entanglement. We indicate how these criteria bear 
on the experimental detection of quantum entanglement.
\end{abstract}

\end{frontmatter}

\section{Introduction}
\label{intro}
The phenomenon of quantum entanglement lies at the heart of quantum mechanics.
And what lies at the heart of quantum mechanics, may lie at the heart of 
a future technology. It is not surprising then that over the last 5 years 
a theory of quantum entanglement has started to emerge that tries to 
capture, quantify and assess the power of quantum entanglement.

First, it was by the protocol of quantum teleportation \cite{tele} that quantum entanglement was introduced as a resource in quantum communication: it has become a rule of quantum communication law that one bit of entanglement (1 ebit) enables 1 unknown qubit to be sent by means of 2 classical bits. 

But it has been realized over the last year that quantum entanglement
is not only a fundamental resource in quantum communication, but can also 
be viewed as a resource in quantum computation. Gottesman and Chuang \cite{gott&chuang} have shown that it is possible to perform universal quantum computation, by starting with three-party entangled GHZ states and subsequently 
performing single qubit operations and measurements in the Bell basis. 
In the linear optics quantum computation proposal by Knill, Laflamme and Milburn \cite{klm} the quantum gate that lies beyond the capabilities of linear optics, can in fact be implemented by the creation of a multipartite entangled state.
Quantum entanglement also lies at the core of the quantum 
computation proposal by Rausschendorf and Briegel \cite{rausbrieg}. In this proposal 
the authors show that universal quantum computation is possible by means of a 
series of single qubit measurements that are performed 
on an initial state which is a certain highly entangled 'cluster' 
state. The entangled state functions as a substrate on which 
the quantum computation takes place.

%The advantage of the division of quantum computation labor into 
%the creation of hard-to-make initial entangled states and the subsequent 
%performance of 'easy' gates, is clear: when our procedure to 
%build the desired initial states fails, we can abort and retry, 
%whereas errors in the middle of the computation need to be dealt 
%with by full-scale error correction. 

In this article we review the progress that has been made in establishing 
one of the cornerstones of the theory of quantum entanglement, namely
the development of criteria for entanglement and separability, both in the 
bipartite as well as the multipartite setting. In the last section of this 
paper we will consider how entanglement witnesses can be used in deciding by experiment whether a quantum state is entangled. 
We will not discuss the topic of entanglement measures, which can be viewed as 
a subject complementary to the one which we consider in this review article.
We would like to refer the reader to Ref. \cite{review:horodecki} for 
a more comprehensive overview on bipartite quantum entanglement.

We will write $X$, $Y$ and $Z$ for the three Pauli matrices. A positive 
semidefinite operator $A$ with nonnegative eigenvalues is denoted 
as $A \geq 0$. A $n$-dimensional Hilbert space is denoted as ${\mathcal H}_n$,
and operators on this space ($n \times n$ matrices) $\in B({\mathcal H}_n)$. 
Furthermore the class of quantum operations which are constructed by Local 
Operations supplemented by Classical Communication is sometimes abbreviated as LOCC.

%We will not discuss the 
%more extensive issue of entanglement measures
%For an extensive review about bipartite quantum entanglent, we refer to
%\cite{horodecki_review}.

\subsection{What is Quantum Entanglement}

A bipartite pure quantum state $\ket{\psi} \in {\mathcal H}_A \otimes {\mathcal H}_B$ 
is called entangled when it cannot be written as $\ket{\psi}=\ket{\psi_A} \otimes \ket{\psi_B}$ for some $\ket{\psi_A} \in {\mathcal H}_A$ and $\ket{\psi_B} \in {\mathcal H}_B$. A mixed state or density matrix $\rho$, which is a positive semidefinite operator on the space ${\mathcal H}_A \otimes {\mathcal H}_B$, is called entangled when it 
cannot be written in the following form 
\beq
\rho=\sum_i p_i \ket{\psi_i^A}\bra{\psi_i^A} \otimes \ket{\psi_i^B}\bra{\psi_i^B},
\label{sepdef}
\eeq
for some set of states $\ket{\psi_i^A} \in {\mathcal H}_A$, $\ket{\psi_i^B} \in {\mathcal H}_B$ and $p_i \geq 0$. If the density matrix $\rho$ can be written in the form of Eq. (\ref{sepdef}), then the density matrix $\rho$ is called separable. 

%\footnote{Rudolph \cite{rudolph_mess} has given an alternative representation %of a necessary and sufficient condition for separability. Consider the followi%ng norm 
%\beq
%||\rho||_{\gamma}=\inf\{\sum_{i} ||X_i||_1 ||Y_i||_1 \,|\,\rho=\sum_i X_i \oti%mes Y_i \},
%\eeq
%where $||X_i||_1=\sqrt{{\rm Tr}(X_i^{\dagger} X_i)}$. Since $||A||_1 \geq {\rm% Tr}(A)$, 
%we have that $||\rho_{\gamma}|| \geq {\rm Tr}(\rho)=1$. When $\rho$ is separab%le,
%%there exists a decomposition as in Eq. (\ref{sepdef}), from which it follows %that% 
%$||\rho||_{\gamma}=1$. On the other hand, if $\rho$ is not separable, then 
%it is necessary to use operators $X_i \not \geq 0$ (or $Y_i \not \geq 0$) 
%with the property that $||X_i||_1 > {\rm Tr}(X_i)$ in all decompositions of $\%rho$ and therefore $||\rho||_{\gamma} > 1$.} 
The distinction between separable states and entangled states has an operational meaning in the following sense. 
A source (or black box) produces a mixed state $\rho$; the mixedness 
could come when aside from systems $A$ and $B$ there are 
additional degrees of freedom to which we have no access. If $\rho$ 
is an entangled density matrix, then some coherent interaction must have taken place between A and B. If $\rho$ is separable, then no guarantee exists
of whether the interaction in the black box was coherent or not.

Consider a multipartite system with parties labeled by $A_1,\ldots A_n$.
The density matrix is called separable when no entanglement exists between 
the parties, i.e. $\rho=\sum_i p_i \ket{\psi_i^{A_1}}\bra{\psi_i^{A_1}} \otimes\ldots \otimes \ket{\psi_i^{A_n}}\bra{\psi_i^{A_n}}$. For the various degrees and forms of quantum entanglement that can exist among parties, we will consider
specific classes of states in Section \ref{multipart}.

A note of caution about how to interpret the state 
of a physical system in terms of quantum entanglement may be in place here. The previous standard 
definitions of quantum entanglement tacitly assume that (1) every state in the bi- or multipartite Hilbert space is in principle available as a physical state and (2) local (involving single tensor factors) as well as global quantum operations, measurements and unitary transformations, can be performed on the Hilbert space. In this respect the wavefunction
of two identical bosons $\Psi(x_1,x_2)=\psi(x_1)\otimes \psi(x_2) +\psi(x_2) \otimes \psi(x_1)$ cannot be called entangled, since it falls short of these 
criteria. Understandably, when considering more complex physical 
systems, the dividing line between what is entangled and what is 
not entangled, may become somewhat fuzzy. The guideline in deciding these matters, I believe, should be the question: ``Do we have an operational form of quantum entanglement? What resource does the particular state constitute in quantum communication and computation?'' In Ref. \cite{fermions_ent} for example the authors consider the entanglement that can exist in 2-fermion systems.

%\cite{fermions_ent}
%Ultimately, it is what we go do with quantum states, that is interesting, and 
%not how we define quantum entanglement.

%These observations are also pertinent in the case that bosonic, fermionic 
%or even anyonic degrees of freedom emerge as quasi-particles from an 
%underlying theory in which a local Hamiltonian acts on a tensorproduct Hilbert% space.  From the point of view of the local underlying structure, these 
%excitations may be highly entangled (see for example the topological 
%quantum computation \cite{kitaev:ft})

%As an example we can consider Cooper pairs, pairs of electrons, which 
%form in low-$T_c$ superconductors. The electron pair which have opposite 
%momenta, is in a 
%a spin singlet state $\frac{1}{\sqrt{2}}(\ket{\uparrow}_{\vec{k}} \otimes \ket%{\downarrow}_{-\vec{k}}-\ket{\downarrow}_{\vec{k}} \otimes \ket{\uparrow}_{-\v%ec{k}})$, so its total spin is zero which allows the pair to behave as boson.
%Since the pair is a boson, we can label a Hilbert space by the momentum 
%$\ket{k}$ and create excitations. Now where is the quantum entanglement?

\section{Bipartite Criteria}

\subsection{Bell Inequalities}
Historically one can say that the first separability criterion was formulated by 
John Bell \cite{bell}. Bell's intention however was not to establish a separability 
criterion, but to evaluate the power of local hidden variable
theories in describing local measurement outcomes on quantum mechanical states.

His inequality, and similar inequalities such as the CHSH inequality \cite{chsh} found later, is obeyed by any local hidden variable theory, whereas the correlations in measurement outcomes on, for example, the singlet state $\ket{\Psi^-}=\frac{1}{\sqrt{2}}(\ket{01}-\ket{10})$ violate the inequality. Furthermore the outcomes of local measurements on any separable density matrix can be simulated by a local hidden variable 
theory. This can be easily understood from Figure \ref{fig1} which 
gives an idea of the workings of a local hidden variable theory.
For every pure entangled state there exists a Bell inequality that is 
violated \cite{peresbook} and therefore there exists a series of measurements
and outcomes through which we can ascertain that our state is entangled.

\begin{figure}[htb]
\epsfxsize=6cm
% input lhv.gif
%\epsffile{lhv.eps}
\vspace{9cm}
\caption{Local Hidden Variable Theories: Alice and Bob each have an arbitrarily powerful machine in their lab which 
takes as input the description of their local measurements ${\mathcal M}_A$ 
and ${\mathcal M}_B$ and a description of the state $\rho$ on which the 
measurement will take place. Inside their machine may be a random shared 
bit string $r$ of arbitrary length. The output of the machines is supposed 
to statistically simulate the outputs of the real measurements ${\mathcal M}_A$ 
and ${\mathcal M}_B$ that were performed on the state $\rho$, in the 
sense that the joint and marginal probabilities for various outcomes and 
choices of measurements are identical to the real measurement on $\rho$.}
\label{fig1}
\end{figure}

\begin{figure}[htb]
\epsfxsize=10cm
% input locc.gif
\vspace{9cm}
\caption{Local Operations and Classical Communication: at each round Alice's
 (Bob's) local actions may depend on Bob's (Alice's) previous actions and 
outcomes.}
\label{fig1b}
\end{figure}

The weakness of Bell inequalities as criteria for entanglement or 
separability, lies in the fact that it is not known whether violations 
exist for many entangled mixed states. For example, it has been shown 
that for a special class of mixed states, so called PPT bound entangled 
states (see Section \ref{ew}), all CHSH-inequalities are obeyed \cite{werner:wolf}.
If we loosen the rules of the game and allow preprocessing of our 
state $\rho$ or many copies of our state $\rho^{\otimes n}$ by means of 
LOCC, then a much larger class of states $\rho$ will violate a Bell inequality. We demand that only local operations and classical communication enter in this game, since these are the operations which cannot increase the quantum entanglement in a state. This class of operations, crucial in the theory of quantum entanglement, is graphically depicted in Fig. \ref{fig1b}. All density matrices which are 
distillable (see Section \ref{dist_locc}) will then violate a Bell inequality in this manner.

\subsection{Entanglement Witnesses and Positive Linear Maps}
\label{ew}

The framework of Bell inequalities fits in a larger scheme of entanglement witnesses. In fact, each Bell inequality can be viewed as 
 a particular example of an entanglement witness \cite{terhalbell}. A prime
example is the operator form of the CHSH inequality: the Bell-CHSH operator ($\vec{a},\vec{a'},\vec{b},\vec{b'}$ are unit-vectors) reads
\beq
{\mathcal B}=\vec{a}\cdot \vec{\sigma}\otimes (\vec{b}+\vec{b'})\cdot \vec{\sigma}+
\vec{a'}\cdot \vec{\sigma}\otimes (\vec{b}-\vec{b'})\cdot \vec{\sigma}.
\label{bellchsh}
\eeq
The expectation value of ${\mathcal B}$ with respect to all separable states 
\beq
{\rm Tr}\, {\mathcal B} \rho_{sep} \leq 2,
\eeq
whereas ${\rm Tr}\, {\mathcal B} \rho$ can exceed this value for an entangled state $\rho$. 
The operator $2{\bf 1}-{\mathcal B}$ is an example of an entanglement witness. Even though there does not necessarily exist a Bell inequality for every entangled state (for certain Werner states, for example, there is no single copy violation of a Bell inequality \cite{werner:lhv}), there does exist a witness for every entangled state. This is the content of the following theorem: 

\begin{thm}(Horodecki \cite{nec_horo})
A density matrix $\rho$ on ${\mathcal H}_A \otimes {\mathcal H}_B$ is 
entangled if and only if there exists a Hermitian matrix $H=H^{\dagger}$,
an entanglement witness, such that 
\beq
{\rm Tr}\, H \rho < 0,
\label{negexp}
\eeq
and for all separable states $\rho_{sep}$,
\beq
{\rm Tr}\,H \rho_{sep} \geq 0.
\label{witpos}
\eeq
\label{ewtheo}
\end{thm}

Thus by a measurement of the entanglement witness observable $H$ we will be able to decide whether a particular state is entangled, since when we find a negative expectation value for $H$, we must conclude that the state cannot be 
separable. 

There exists a direct relation between entanglement witnesses and positive 
linear maps which are not completely positive \cite{jamiolkowski}. A linear map ${\mathcal L}\colon B({\mathcal H}_n) \rightarrow
B({\mathcal H}_m)$ is called positive, when it maps all $X \geq 0$ onto 
${\mathcal L}(X) \geq 0$. The map ${\mathcal L}$ is completely positive if 
and only if ${\bf 1}_n \otimes {\mathcal L}$ is a positive map. The most 
famous and physically relevant example of a positive map is matrix transposition $T$ in a chosen basis. The map $T$ is not completely positive, as can be 
illustrated by applying it on half of a (unnormalized) maximally entangled state:
\bea
\lefteqn{({\bf 1} \otimes T)[\ket{00}+\ket{11})(\bra{00}+\bra{11}]=} \hspace{2cm}\nonumber \\
& & \ket{00}\bra{00}+\ket{11}\bra{11}+\ket{01}\bra{10}+\ket{10}\bra{01}.
\eea
The resulting operator has an eigenvector $\ket{01}-\ket{10}$ with negative 
eigenvalue $-1$. It was noted by Peres \cite{Asher96} that applying ${\bf 1} \otimes T$ on a separable density matrix always gives another density matrix
\bea
\lefteqn{({\bf 1} \otimes T)(\sum_i p_i \ket{\psi_i^A}\bra{\psi_i^A}\otimes \ket{\psi_i^B}\bra{\psi_i^B})=} \hspace{3.3cm}\nonumber \\
& & \sum_i p_i \ket{\psi_i^A}\bra{\psi_i^A}\otimes \ket{{\psi_i^B}^*}\bra{{\psi_i^B}^*} \geq 0,
\eea
and therefore the condition $({\bf 1}\otimes T)(\rho) \geq 0$, sometimes called the Peres-Horodecki criterion, constitutes a separability criterion.

The relation between entanglement witnesses and positive linear map is the 
following. We take a maximally entangled state in ${\mathcal H}_n \otimes {\mathcal H}_n$, for example the state $\ket{\Psi^+}=\sum_{i=1}^n \ket{i,i}$. The 
Hermitian operator $H \in B({\mathcal H}_n \otimes {\mathcal H}_m)$ defined by 
\beq
H=({\bf 1}\otimes {\mathcal L})(\ket{\Psi^+}\bra{\Psi^+}),
\eeq
has the property of Eq. (\ref{witpos}) if and only if ${\mathcal L}\colon B({\mathcal H}_n) \rightarrow B({\mathcal H}_m)$ is a
positive map. Furthermore, $H$ is an entanglement witness, as in Eq. (\ref{negexp}), if and only if ${\mathcal L}$ is not a completely positive map.

The theory of positive maps which are not completely positive has not been 
completely developed. What is known is that in spaces such as 
${\mathcal H}_2 \otimes {\mathcal H}_2$ and ${\mathcal H}_2 \otimes {\mathcal H}_3$, all positive maps ${\mathcal L}$ relate to the matrix transposition 
map $T$, i.e. they can all be written as 
\beq
{\mathcal L}={\mathcal S}_1+
{\mathcal S}_2 \circ T,
\label{decomp}
\eeq
where ${\mathcal S}_1$ and ${\mathcal S}_2$ are completely positive maps and $T$ is matrix transposition in any chosen basis \cite{woronowicz}. This implies that for these small dimensions, ${\mathcal H}_2 \otimes {\mathcal H}_2$ or ${\mathcal H}_2 \otimes {\mathcal H}_3$, the entanglement witnesses 
are of the form
\beq
H=P+({\bf 1} \otimes T)(Q),
\label{genform22}
\eeq
where $Q \geq 0, P \geq 0$ and $T$ is matrix transposition in any chosen 
basis.

In higher dimensions the situation is more involved. There do exist 
positive maps which are not {\em decomposable}, meaning that they are 
not of the form of Eq. (\ref{decomp}). A consequence is that 
in higher dimensions, there exist {\em entangled} states which satisfy the 
Peres-Horodecki criterion, i.e. 
$({\bf 1}\otimes T)(\rho) \geq 0$ where $T$ is matrix transposition in any basis. These are the bound entangled density 
matrices with PPT (Positive Partial Transposition) (see the next Section). The first example of an indecomposable positive map in ${\mathcal H}_3 \otimes {\mathcal H_3}$ was found by M.-D. Choi \cite{choi:biquad}. The first method of constructing some indecomposable entanglement witnesses in arbitrary high dimensions was presented in Ref. \cite{terhalposmap}. In Refs. \cite{lewpos1} and \cite{lewpos2} this construction was generalized with the following consequences. It was shown in Ref. \cite{lewpos1} that every indecomposable entanglement witness is of the form 
\beq
H=P+({\bf 1}\otimes T)(Q)-\eps {\bf 1},
\label{indecomp}
\eeq
where, in order to ensure that $H$ has the property of Eq. (\ref{witpos}), we have 
\beq
0 < \eps \leq \inf_{\psi_A,\psi_B} \bra{\psi_A,\psi_B}\, (P+({\bf 1}\otimes T)(Q))\,\ket{\psi_A,\psi_B}.
\eeq
Here $P \geq 0$, $Q \geq 0$ and they are such that ${\rm Tr}\, P \delta=0$ and ${\rm Tr}\, Q \,({\bf 1}\otimes T)(\delta)=0$ for an 'edge' state $\delta$. 
An edge state $\delta$ is a bound entangled PPT state which has the property that for all $\eps > 0$ and all product states $\ket{\psi_A,\psi_B}$, $\delta-\eps \ket{\psi_A,\psi_B}\bra{\psi_A,\psi_B}$ is not positive or does not have PPT.
As such, the edge states are on the boundary of the closed set of entangled PTT states.  The entanglement witness $H$ in Eq. (\ref{indecomp}) detects
the entanglement of the edge state $\delta$. 

The entanglement of bound entangled PPT states $\rho$ in the interior of the set of PPT states is also detected by these witnesses \cite{lewpos1}. By choosing 
${\rm Tr}(P+({\bf 1}\otimes T)(Q))\rho=0$ we ensure that the indecomposable
witness $H$ in Eq. (\ref{indecomp}) detects the entanglement in $\rho$, 
i.e. ${\rm Tr} H \rho < -\eps  < 0$.

Thus, given an edge state $\delta$, its entanglement witness can be 
determined. A complete characterization of these edge states is still an 
open question; in Ref. \cite{lewpos1} they are shown to be based on 
pairs of subspaces ${\mathcal H}_1$ and ${\mathcal H}_2$ such that 
(1) for every product state $\ket{\psi_A,\psi_B} \in {\mathcal H}_1$, 
$\ket{\psi_A, \psi_B^*} \not \in {\mathcal H}_2$ and (2) the rank of 
${\rm Tr}_i P_{{\mathcal H}_1}$ is equal to the rank of ${\rm Tr}_i P_{{\mathcal H}_2}$ for $i=A,B$, where $P$ is the projector onto the subspace. For the choice ${\mathcal H}_1={\mathcal H}_2$, an example of such a subspace (containing, in this case, no product vectors) is the space orthogonal to an unextendible product basis \cite{upb1}. It would be very interesting
to find a method for constructing such subspaces in general.

Indecomposable positive maps are highly nontrivial objects. As was noted by M.-D. Choi they provide special counterexamples to Hilbert's 17th problem. 
In Appendix \ref{hilbert17} we will review this connection.

%It is interesting that the existence of 
%PPT bound entanglement is rooted so deeply in mathematics.

\subsection{Operational criteria: LOCC and distillability}
\label{dist_locc}

The resource view of quantum entanglement emerged with the discovery 
of quantum teleportation \cite{tele}. A natural next question, which 
was asked and partially answered by Bennett {\em et al.} in Ref. \cite{bdsw} 
is, in what sense does mixed state entanglement enables quantum data transmission? 
This line of study, which is still ongoing, has 
led to operational criteria for 'useful' entanglement. We can classify 
entangled states in terms of their distillability.
Distillation of a mixed state $\rho$ is a process which is implemented by 
LOCC on a large set of copies of the 
state $\rho^{\otimes n}$. The process outputs a smaller set of states $\sigma^{\otimes k}$ on a space of dimension $2^k \times 2^k$ such that $\bra{{\Psi^-}^{\otimes k}}\, \sigma^{\otimes k}\, \ket{{\Psi^-}^{\otimes k}} \rightarrow 1$ when 
$n \rightarrow \infty$. Here $\ket{\Psi^-}$ is the singlet state. The asymptotic fraction $\frac{k}{n}$ is called the distillable entanglement $D$ of the density matrix $\rho$. It has been shown that all entangled density matrices in 
${\mathcal H}_2 \otimes {\mathcal H}_2$ are distillable \cite{dist2d}. In fact, a generalization of this result exists: all states which violate the so called
{\em reduction criterion} are known to be distillable \cite{filterhor}. The 
reduction criterion is violated for a density matrix $\rho$ when either 
\bea
{\bf 1}\otimes \rho_B-\rho \not \geq 0 & \mbox{ or } \rho_A \otimes {\bf 1}-\rho \not \geq 0.
\label{red}
\eea
%It is not hard to see that a separable density matrix $\rho=\sum_i p_i \ket{\p%si_i,\phi_i}\bra{\psi_i,\phi_i}$ obeys the reduction criterion: 
It is noteworthy that satisfaction of the reduction criterion is identical \cite{filterhor} to positivity under partial application of the decomposable positive map \linebreak ${\mathcal L}(X)={\bf 1}\,{\rm Tr}X-X$, since $({\bf 1}\otimes {\mathcal L})(\rho)=\rho_A \otimes {\bf 1}-\rho$. Since the map ${\mathcal L}$ is decomposable, Eq. (\ref{decomp}), it follows that every state which satisfies the Peres-Horodecki criterion will also satisfy the reduction criterion.

The class of density matrices which are not distillable are called {\em bound entangled} density matrices. At least 
one group of density matrices exists for which it is possible to rigorously 
prove that they are bound entangled density matrices. These are entangled density matrices $\rho$ which do not violate the Peres-Horodecki criterion. This 
criterion is important in the theory of quantum entanglement, since it has been shown \cite{pptnodist} that nonviolation of the Peres-Horodecki criterion is preserved under LOCC. This is not hard to see, when we consider a larger class of quantum operations, the separable superoperators, of which the LOCC actions are a subset. A separable
superoperator or measurement \cite{rainsdist} has operation elements that are of separable form $\{A_i \otimes B_i, \sum_i A_i^{\dagger}A_i \otimes B_i^{\dagger} B_i={\bf 1}\}$. Given is 
a density matrix $\rho$ such that $({\bf 1}\otimes T)(\rho) \geq 0$. For every $i$, we can write 
\beq
({\bf 1}\otimes T)[A_i \otimes B_i \rho A_i^{\dagger} \otimes B_i^{\dagger}]=
(A_i \otimes B_i^*) ({\bf 1}\otimes T)(\rho) (A_i^{\dagger}\otimes B_i^T) \geq 0,
\eeq
under the premise that $\rho$ satisfies the Peres-Horodecki criterion. Thus under the action of a separable superoperator or measurement, PPTness is preserved. Since the output of a distillation process (arbitrarily good approximations to singlet states) does violate the Peres-Horodecki criterion, it must follow that 
entangled states with the PPT property cannot be distilled. The first bound 
entangled states were found by P. Horodecki \cite{bepawel}. In Refs. \cite{upb1} 
and \cite{upb2} constructions were given for classes of bound entangled states based on unextendible product bases.

A second class of states has been conjectured to be nondistillable \linebreak \cite{nptnond1,nptnond2}. Examples of these states are particular Werner states in 
${\mathcal H}_n \otimes {\mathcal H}_n$, of the form 
\beq
\rho_{\lambda}=\frac{1}{\lambda(n^2-1)-1}\left(\lambda {\bf 1}-(\lambda+1)({\bf 1}\otimes T)(\ket{\Psi^+}\bra{\Psi^+})\right),
\label{werners}
\eeq
where $\ket{\Psi^+}=\frac{1}{\sqrt{n}}\sum_{i=1}^n \ket{i,i}$. For all finite 
$\lambda \geq 0$, these Werner states violate the Peres-Horodecki criterion. The states are 
conjectured to be nondistillable for $\lambda \in [\frac{2}{n-2},\infty)$.

The structure among the set of (conjectured) nondistillable states is by 
itself fairly complex. It has been shown recently that the tensorproduct of 
two nondistillable states, one of the PPT kind and one nondistillable Werner state, Eq. (\ref{werners}), in ${\mathcal H}_3 \otimes {\mathcal H}_3$, can be a distillable state \cite{sst:bip}. Furthermore, it is not clear what classes
of bound entangled states are asymptotically interconvertible by local actions and classical communication. In order to investigate
this 'fine-structure' we may need to look for positive maps ${\mathcal L}$ for 
which ``positivity under partial application of ${\mathcal L}$'' is preserved 
under LOCC. 

\subsection{Functional Separability Criteria} 
\label{sepcrit}

Even though entanglement witnesses completely characterize the set 
of separable states, they do not provide a simple computational method 
(except for the Peres-Horodecki condition) for deciding whether a density matrix is 
entangled. It is desirable to have alternative or additional criteria that 
may help us in deciding this and characterizing separable versus 
entangled states. The criteria that we will discuss in this Section are all obeyed by separable density matrices, so that in case of a violation we 
know that $\rho$ is entangled. 

The first criterion is a combination of the Peres-Horodecki criterion and a check 
on the rank of the density matrix. When we find that $\rho$ has PPT,
$\rho$ may either be separable or bound entangled. In Ref. \cite{berank} 
it was proved that a density matrix $\rho \in B({\mathcal H}_n \otimes {\mathcal H}_m)$ with PPT and a rank which is smaller than or equal to $\max(n,m)$ is separable. For PPT density matrices for which the sum of the rank of $\rho$ and $({\bf 1}\otimes T)(\rho)$ is smaller than or equal to $2mn-m-n+2$, the authors
in Ref. \cite{berank} provide an algorithm for checking whether $\rho$ is separable.

It can be shown that from a non-violation of the reduction criterion several 
other criteria can be derived:

\begin{lem}
For all separable states the reduction criterion is not violated and this implies that if $\rho$ is separable,
\bea
S_{\alpha}(\rho_A) \leq S_{\alpha}(\rho) & \mbox{ and } 
S_{\alpha}(\rho_B) \leq S_{\alpha}(\rho).
\label{entr}  
\eea
for $\alpha=0$, $\alpha \in [1,2]$ and $\alpha=\infty$ where $S_{\alpha}(\rho)$ for $0 < \alpha < \infty$ ($\alpha \neq 1$) is the quantum Renyi entropy
\beq
S_{\alpha}(\rho)=\frac{1}{1-\alpha}\log {\rm Tr}\,\rho^{\alpha}.
\eeq
For $\alpha=0$, we have $S_0(\rho)=\log R(\rho)$ where $R(\rho)$ is the rank of $\rho$, $\lim_{\alpha \rightarrow 1} S_{\alpha}=S(\rho)$ where $S$ is the von Neumann entropy $S(\rho)=-{\rm Tr}\, \rho \log \rho$ and for $\alpha=\infty$ we have $S_{\infty}=-\log ||\rho||$.
\label{entrop}
\end{lem}

The case $\alpha=\infty$ was proved in Ref. \cite{filterhor} and the case $\alpha=0$ was proved in Ref. \cite{norank2}. The cases $\alpha \rightarrow 1$ and $\alpha=2$ can be derived from the reduction criterion \cite{privhor}. For $\alpha \rightarrow 1$, we use the operator-monotoniticity \cite{majando} of the $\log$ function to infer that
\beq
\log \rho_A \otimes {\bf 1} \geq \log \rho.
\eeq
Now we use that when $X \geq 0$, ${\rm Tr}\, \rho X \geq 0$ for all $\rho \geq 0$. Thus, multiplying with $\rho$ on both sides and subsequently taking the trace gives the desired entropy inequality for $\alpha \rightarrow 1$. 
If $\rho_A \otimes {\bf 1} \geq \rho$, then $(\rho_A \otimes {\bf 1})^{\delta}
\geq \rho^{\delta}$ for $\delta \in (0,1]$, since $t^{\delta}$ is operator-monotone for $\delta$ in this interval \cite{majando}. If we multiply the inequality with $\delta$ by $\rho$ on both sides and trace, we get 
\beq
{\rm Tr}_A \rho_A^{1+\delta} \geq {\rm Tr} \rho^{1+\delta}, 
\eeq 
which, after taking logarithms on both sides, proves the result for $\alpha \in (1,2]$. 
 
In Ref. \cite{cag} the conditional quantum operator was defined  
\beq
\rho_{A|B}=\exp[\log \rho-\log {\bf 1}_A \otimes \rho_B].
\eeq
With this definition the conditional entropy $S(A|B)=-{\rm Tr}\rho \log 
\rho_{A|B}$ is the difference between the total entropy of the state 
$S(\rho)$ and the local entropy $S(\rho_B)$. Lemma \ref{entrop} states 
that for all states obeying the reduction criterion (separable states 
and at least all nondistillable states) such a conditional entropy 
(and similarly some $\alpha$-entropic extensions) is nonnegative.

Nielsen and Kempe \cite{niel_kemp} recently found a different separability criterion
\begin{lem} \cite{niel_kemp} For all separable density matrices $\rho$
\bea
\vec{\lambda}_{\rho_A} \succ \vec{\lambda}_{\rho} & \mbox{ and  }\vec{\lambda}_{\rho_B} \succ \vec{\lambda}_{\rho},
\eea
where $\vec{\lambda}_{\sigma}$ is the ordered vector of eigenvalues of the
density matrix $\sigma$. Here the symbol $\succ$ means majorization, i.e. 
$\vec{\lambda} \succ \vec{\mu}$, when $\sum_{i=1}^k \lambda_i \geq \sum_{i=1}^k
\mu_i$ for all $k$.
\label{nielkemp}
\end{lem}

Similar to the entropic criteria given above, this majorization 
criterion affirms the intuition that separable density matrices are globally at least as mixed as locally. A new corollary of the majorization criterion are the entropic inequalities in Eq. (\ref{entr}) for $\alpha \in [0,1]$. This follows from the fact that the quantum Renyi entropy $S_{\alpha}(\rho)$ is a {\em concave} function of the probabilities $\vec{\lambda}_{\rho}$ for all $\alpha \in [0,1]$. In Ref. \cite{wehrl} it is claimed, in fact, that the majorization 
condition (Uhlmann's relation) implies the entropic inequalities for all
$\alpha \geq 0$.

%It may be conjectured that the entropic inequalities hold for separable states
%for all $\alpha \geq 0$.

The two criteria, the reduction criterion of Eq. (\ref{red}) and Lemma \ref{nielkemp} are strikingly similar. There exist states however for which the reduction criterion is violated whereas the majorization criterion is satisfied; this is the example of a 2 qubit entangled state in Ref. \cite{niel_kemp}. The reduction criterion is violated for this state, since in ${\mathcal H}_2 \otimes {\mathcal H}_2$ the reduction criterion is equivalent to the Peres-Horodecki criterion, and all entangled 2-qubit states violate this criterion. The conjectured nondistillable Werner states, Eq. (\ref{werners}), obey both the reduction as well as the majorization criterion. 
It is possible, but unproven, that any state which satisfies the reduction 
criterion also satisfies the majorization criterion.

Unfortunately, non-violation of the reduction criterion and non-violation of the majorization criterion are {\em not} properties that are preserved under local actions and classical communications.
To take an example, consider the following density matrix 
$\sigma={\bf 1}_A/d \otimes \ket{\Psi^+}\bra{\Psi^+} \otimes {\bf 1}_B/d$ where ${\bf 1}_A/d$ (${\bf 1}_B/d$) is a density matrix for Alice (Bob), $\ket{\Psi^+}$ is a maximally entangled state in ${\mathcal H}_n \otimes {\mathcal H}_n$ and $d$ is large. The state $P_+=\ket{\Psi^+}\bra{\Psi^+}$ violates the reduction criterion
\beq
{\bf 1}/n-P_+\not \geq 0,
\eeq
We can always choose $d > n$ large enough such that both the reduction criteria are satisfied for $\sigma$: 
\bea
\lefteqn{{\bf 1}_A \otimes [{\bf 1}_A \otimes {\bf 1}_B/n-P_+/d]\otimes {\bf 1}_B \geq 0, \mbox{   and   }}\hspace{4cm} \nonumber \\
& & {\bf 1}_A \otimes [{\bf 1}_A/n \otimes {\bf 1}_B-P_+/d] \otimes {\bf 1}_B \geq 0.
\eea
By the local action of tracing over the register with ${\bf 1}_A/d$ and ${\bf 1}_B/d$, we obtain the state $P_+$ which does violate the reduction criterion. Thus there exists states which initially do not violate the reduction criterion, but nonetheless are distillable. The same example can serve
to show that the majorization criterion can be violated {\em only after} some local action. 

In Figure \ref{fig2sep} we have sketched an overview of the known relations between the various bipartite separability criteria.

\begin{figure}[htb]
\begin{center}
\epsfxsize=10cm 
% input sets.gif
\vspace{9cm}
\caption{Relations between various bipartite separability criteria. It is not 
known whether the green (Reduction Criterion) and light blue (Peres-Horodecki) ellipses are contained in the dark blue (Majorization) ellipse. In ${\mathcal H}_2 \otimes {\mathcal H}_2$ all colored ellipses, except the dark blue one, collapse onto each other and mark the separation of the 
set of separable states from the entangled states. } 
\end{center}
\label{fig2sep}
\end{figure}

%The separability criteria that were formulated by Cerf, Adami and Gingrich
% follow from the satisfaction of the reduction criterion

\section{Multipartite Entanglement}
\label{multipart}

Every time we consider a multipartite system in a bipartite fashion,
that is, we split the set of parties in two subsets, and consider the 
entanglement between the subsets, we can apply the criteria that we have 
listed in the previous Section. In this Section we will focus on features
of multipartite quantum entanglement which are special to multipartite quantum entanglement.

\subsection{Violations of Local Realism and Bell inequalities}
\label{bellmulti}

The Greenberger-Horne-Zeilinger state $\ket{GHZ}=\frac{1}{\sqrt{2}}(\ket{000}-\ket{111})$ is an example
of a three party state which violates the predictions of local realism \cite{mermin}. The GHZ-state is an eigenvector of operators $X \otimes Y \otimes Y$, $Y \otimes X \otimes Y$, $Y \otimes Y \otimes X$ with eigenvalues $+1$ and $X \otimes X \otimes X$ with eigenvalue $-1$. These first three operators form the generators of an abelian group $G$ which contains $X \otimes X \otimes X$. 
%Any local 
%operator can be measured without changing the density matrix for the other
%two particles. 
For these four operators, from a measurement of $X$ or $Y$ on two of the spins we can deduce the outcome of the third since $\ket{GHZ}$ is an eigenstate. Therefore,
according to local realism, we may assign values to the local operators $X_i$ and $Y_i$ according to some function $f$
\bea
f\colon X_i \rightarrow \{-1,+1\}\;\; & f\colon Y_i \rightarrow \{-1,+1\},
\eea
where $X_i$ acts on the $i$th particle, obeying the eigenequation constraints.
The function $f$ gives rise to a function $h \colon g \in G \rightarrow \{-1,+1\}$, i.e. $h(X \otimes Y\otimes Y)=f(X) f(Y) f(Y)$. A violation of local realism occurs when it is impossible to construct a local function $f$ (consistent with the eigenvalue equations for the generators) such that $h$ is a {\em group homomorphism}, i.e. $h(g_1 \circ g_2)=h(g_1) \circ h(g_2)$ for all $g_1,g_2 \in G$. In the case of the GHZ-state, the violation comes about by observing that the local assignments (the function f) for the generators $X \otimes Y \otimes Y$, $Y \otimes X \otimes X$ and $X \otimes X \otimes Y$ always give $h(X \otimes X \otimes X)=1$ whereas the 
GHZ-state has eigenvalue $-1$ with respect to $X \otimes X \otimes X$.
 
In Ref. \cite{divperes} violations of local realism were found 
for a special class of multipartite entangled states. These are states used in 
quantum error correction codes based on the stabilizer formalism. The states are by definition the eigenvectors of an Abelian group made from tensor products of Pauli matrices and ${\bf 1}$.

These violations do not yet present us with a separability criterion. The only claim is that if the outcomes of local measurements of $X$ and $Y$ were to correspond {\em exactly} to, say, the outcomes on the GHZ-state, then we 
may conclude that these measurement outcomes cannot be described by a local hidden variable theory. However, to establish a full separability criterion, we would need to analyze what ranges of outcomes could still be reproduced by a separable state or local hidden variable theory and what outcomes cannot. 

The $n$-qubit Bell-Klyshko operator \cite{bech:bell} or Mermin's inequality \cite{mermin_multi} in operator form (see for example Ref. \cite{werner:wolf}) do constitute a separability criterion in this way. The Bell-Klyshko operator can be defined recursively
\beq
{\mathcal B}_n={\mathcal B}_{n-1}\otimes \frac{1}{2}(\vec{a}_n \cdot \vec{\sigma}+\vec{a}'_n \cdot \vec{\sigma})+{\mathcal B}'_{n-1}\otimes \frac{1}{2}(\vec{a}_n \cdot \vec{\sigma}-\vec{a}'_n \cdot \vec{\sigma}),
\eeq
where ${\mathcal B}'_{n-1}={\mathcal B}_{n-1}(\vec{a}_1 \leftrightarrow \vec{a}'_1, \vec{a}_2 \leftrightarrow \vec{a}'_2, \ldots, \vec{a}_{n-1} \leftrightarrow \vec{a}'_{n-1})$ and ${\mathcal B}_2$ is 
the Bell-CHSH operator in Eq. (\ref{bellchsh}). Its expectation value for all separable states is bounded as ${\rm Tr}\, {\mathcal B}_n \rho_{sep} \leq 2$, whereas a value as large as $2^{(n+1)/2}$ can be obtained for the cat state $\ket{GHZ_n}=\frac{1}{\sqrt{2}}(\ket{0^{\otimes n}}+\ket{1^{\otimes n}})$. As in the bipartite case, these operators are examples of entanglement witnesses.

\subsection{Entanglement Witnesses and Linear Maps}
\label{ewmulti}

In Ref. \cite{nsep:horodecki} the notion of entanglement witnesses was 
formally extended to the domain of multipartite quantum systems. The multipartite analog of Theorem \ref{ewtheo} separates the multipartite 
separable states from any entangled state; for every state $\rho$, not 
of the form $\rho=\sum_i p_i \ket{\psi_i^{A_1}}\bra{\psi_i^{A_1}} \otimes\ldots \otimes \ket{\psi_i^{A_n}}\bra{\psi_i^{A_n}}$, there exists a 
Hermitian operator $H$, the entanglement witness, such that 
\beq
{\rm Tr} H \rho_{sep} \geq 0,
\label{seppos}
\eeq
for all separable states $\rho$ and ${\rm Tr}\, H \rho < 0$. Interestingly,
these multipartite witnesses can be shown \cite{nsep:horodecki} to relate to linear maps ${\mathcal L} \colon B({\mathcal H}^2\otimes {\mathcal H}^3 \otimes \ldots \otimes {\mathcal H}^n) \rightarrow B({\mathcal H}^1)$ with the property that for all product vectors $\ket{x_2,\ldots,x_n}$,
\beq
{\mathcal L}(\ket{x_2,\ldots,x_n}\bra{x_2,\ldots,x_n}) \geq 0.
\label{linpartpos}
\eeq
The linear map ${\mathcal L}$ is thus not necessarily positive on $B({\mathcal H}^2\otimes {\mathcal H}^3 \otimes \ldots \otimes {\mathcal H}^n)$; this 
is an important difference with the bipartite case. The 1-1 relation between a 
witness $H$ with the property of Eq. (\ref{seppos}) and the map ${\mathcal L}$ 
with the property of Eq. (\ref{linpartpos}) is the following: 
\beq
H=({\bf 1}_1 \otimes {\mathcal L}^{\dagger})(\ket{\Psi^+}\bra{\Psi^+}),
\eeq
where $\ket{\Psi^+}$ is a maximally entangled state in ${\mathcal H}^1 \otimes {\mathcal H}^1$. Here the Hermitian 
conjugate of ${\mathcal L}$, ${\mathcal L}^{\dagger}$ is defined by the relation ${\rm Tr}\, A^{\dagger} {\mathcal L}(B)={\rm Tr}\,{\mathcal L}^{\dagger}(A^{\dagger}) B$.

In Ref. \cite{janzing:ew} a family of multiqubit entanglement witnesses was presented. For an $n$-partite qubitsystem the authors introduce an averaging observable $\bar{a}=\frac{1}{n}\sum_{i=1}^n a_i$ where $a_i$ is an observable acting on the $i$th factor in ${\mathcal H}^1 \otimes {\mathcal H}^2 \otimes \ldots \otimes {\mathcal H}^n$. The operators $a_i$ are bounded in their norm, $||a_i|| \leq 1$ (here $||.||$ is the standard operator norm). The operator $\bar{a}$ 
could, for example, be a sum of Pauli operators $Z_i$, measuring a mean 
magnetic field in the $z$-direction. It is proved in Ref. \cite{janzing:ew} that for separable density matrices $\rho$ the expectation of the commutator 
\beq
|{\rm Tr}\, \rho [\bar{a},c]|\leq \frac{2}{\sqrt{n}},
\label{expac}
\eeq
for any averaging observable $\bar{a}$ and $c=c^{\dagger}$ with $||c|| \leq 1$. An expectation value such as Eq. (\ref{expac}) can appear in a perturbative 
expansion in linear response theory (see for example Section IV in Ref. \cite{terhal&divincenzo:equilib}). The operator $c$ will then be a nonlocal time-dependent observable, the operator $\bar{a}$ some local mean field observable and $\rho$ can be the equilibrium state of a
physical system. For entangled states the expectation value of the commutator in Eq. (\ref{expac}) can be of order 1 as was shown in Ref. \cite{janzing:ew}.
A consequence is that for such entangled states the region of validity of 
the perturbation theory is much smaller than for separable states. Let us translate the result in the language of entanglement witnesses. The observable
\beq
H=\frac{2}{\sqrt{n}}{\bf 1}- i [\bar{a},c],
\eeq
is a witness in the sense that Eq. (\ref{seppos}) holds. This witness can detect the entanglement in superpositions of macroscopically distinct states, such as the cat $\ket{GHZ_n}$ state. For example (see Ref. \cite{janzing:ew}) when we choose the operators $a_i=\ket{1}\bra{1}_i$ and $c=i[\ket{0^{\otimes n}}\bra{1^{\otimes n}}-\ket{1^{\otimes n}}\bra{0^{\otimes n}}]$, we obtain a witness which has
\beq
\bra{GHZ_n}\, H \ket{GHZ_n}=\frac{2}{\sqrt{n}}- 1, 
\eeq
which is negative when $n > 2$.

\subsection{Incomparable Forms of Pure State Entanglement}

The possibility for extraction of bipartite pure state entanglement from mixed state entanglement by LOCC is captured by the notion of distillation, which we discussed in Section \ref{dist_locc}. The definition of bipartite 
distillation does not depend on the form, --maximally or partially entangled states--, of the final pure state entanglement, since the
{\em Asymptotic Interconversion Theorem} for bipartite entanglement \cite{bbps}
says that all bipartite pure state entanglement is interconvertible
by LOCC in the asymptotic limit (when we have many copies of a state).
For multipartite entanglement no such theorem exists. The exploration of 
interconvertibility of multipartite entanglement was initiated in Ref. \cite{multibenn}. It was found for example that 3 EPR pairs $(\ket{00}+\ket{11})^{\otimes 3}$ shared among three parties were not exactly convertible by LOCC to 2 GHZ states $(\ket{000}+\ket{111})^{\otimes 2}$. In Ref. \cite{linden:rev} this result was considerably strengthened by showing that $3n$ EPR pairs are not interconvertible to $2n$ GHZ states even in the asymptotic limit $n \rightarrow \infty$. As such these states form the building blocks of an MREGS, a Minimal Reversible Entanglement Generating Set \cite{multibenn}, with which all tripartite 
entanglement can reversibly be created. It is an open question whether
a third type of state, the $W$-state $\ket{001}+\ket{010}+\ket{100}$ should be added to the tripartite MREGS, in other words whether the $W$-state is asymptotically interconvertible to a supply of EPR pairs and GHZ states (see Refs. \cite{durvidalcirac} and \cite{galvao} for indications that this may not be the case). These results show that intrinsically {\em multipartite} forms of pure state entanglement exist.

\subsection{Bound Entanglement}
\label{bemulti}

One of the best illustrations of the phenomenon of intrinsic mixed state 'multipartiteness'
was given in Ref. \cite{upb1}. It is an example of a tripartite mixed state in ${\mathcal H}_2 \otimes {\mathcal H}_2 \otimes {\mathcal H}_2$ which is 
separable over all bipartite cuts of the three parties, while at the same
time the state is entangled. Let $\ket{v_1}=\ket{000}$, $\ket{v_2}=\ket{-+1}$,$\ket{v_3}=\ket{+1-}$ and $\ket{v_4}=\ket{1-+}$, where $\ket{\pm}=\frac{1}{\sqrt{2}}(\ket{0}\pm\ket{1})$. The state is
\beq
\rho_{\bf Shifts}={\bf 1}-\sum_{i=1}^4 \ket{v_i}\bra{v_i}.
\label{shiftbe}
\eeq
No product state state exists in the range of $\rho_{\bf Shifts}$, since the vectors \linebreak
$\{\ket{v_1},\ket{v_2},\ket{v_3},\ket{v_4}\}$ form an unextendible product basis
\cite{upb1}. At the same time, the product basis can be completed with 
vectors which are separable over a bipartite cut, which results in $\rho$ 
being separable over this cut. 
The density matrix $\rho_{\bf Shifts}$ in Eq. (\ref{shiftbe}) is also an example of a 
{\em bound entangled} state in the multipartite setting; if entanglement 
could be distilled from $\rho_{\bf Shifts}^{\otimes n}$ by local actions and classical
communications of the three parties, then entanglement would be created 
over some bipartite cut, which is forbidden since $\rho$ is separable
over all cuts.

The phenomenon of multipartite bound entanglement is more general than 
this. It can be argued \cite{smolin} that any multipartite density matrix $\rho$ for parties 
$A_1,\ldots, A_k$ is bound entangled when for all pairs of parties 
$(A_i,A_j)$ there exists a cut (a bipartition) where $A_i$ and $A_j$ are 
in different sets of parties, such that $\rho$ is either 
separable or PPT over this cut. This follows from the fact that these 
separability and PTT properties are preserved under LOCC and the fact that 
any multipartite pure entangled state is entangled over some bipartition.
In Ref. \cite{durcirac} and Ref. \cite{sst:act} multiqubit examples of such bound entangled states were presented. Moreover in Ref. \cite{sst:act} it was 
shown that two $4$-party bound entangled states both distributed among 5 parties, can be distillable. The state $\rho^{ACBD}$ is a mixture of Bell states; with probability $1/4$, A and C share one of the 4 Bell states and B and D share {\em the same} Bell state. The state is symmetric under permutation of parties. Furthermore it is separable over any bipartition into $(2,2)$, but it is entangled for any bipartition into $(3,1)$. The other state $\rho^{ABCE}$ is identical, except that now it is shared among $A$, $B$, $C$ and $E$. The essential feature of these two states taken together, is that 
there exists no bipartition over which $\rho^{ACBD} \otimes \rho^{ABCE}$ 
is separable such that the parties $D$ and $E$ belong to different sets.
This fact makes it possible for entanglement to be distilled between $D$ and 
$E$ and the authors of Ref. \cite{sst:act} show that 1 singlet (1 ebit) can 
be obtained from the two bound entangled states.

%\subsection{Robustness of Entanglement: Clusters, Molecules and Chains}

%The $n$-partite GHZ state $\ket{0^{\otimes n}}+\ket{1^{\otimes n}}$, 
%an equal superposition of two macroscopic states, is a very fragile entangled %state. Tracing over any 1 of the spins leaves us with a separable state betwee%n the other $n-1$ parties. At the same
%time the GHZ-state is a powerful resource for entanglement; an EPR singlet 
%$\ket{\Psi^-}$ can be created between any pair of parties when the other 
%parties perform measurements in $\{\ket{+}\,\ket{-}\}$ basis and communicate
%the outcomes. In Ref. \cite{briegel_pers}, a new type of multipartite pure sta%te entanglement was introduced, which has more of the desirable property of ro%bustness than the GHZ-state, but nonetheless can be used to create ebits betwe%en any two parties by LOCC. For 

%In the entanglement chains introduced by Wootters \cite{woot_echain}
%Molecules by D\"ur

\section{Experimental Issues: Detecting Quantum Entanglement}

In this Section we consider how the criteria that we have discussed in the previous Sections enable us to decide by physical experiment whether the physical state of a given quantum state is entangled. The capacity to build certain
entangled states is one of the basic requirements for making a quantum 
computer and is often used as a benchmark test for the amount of control
and coherence in a particular quantum system, see for example the creation 
of the cat state $\ket{0^{\otimes 7}}+\ket{1^{\otimes 7}}$ in the NMR experiment in Ref. \cite{nmrcat}.

It is desirable that the verification of the entanglement take place with a minimal number of measurements and operations. The first, but inefficient option would be to perform full quantum tomography on the state $\rho$, i.e. determine
all the matrix elements $\rho_{ij}$, after which we may analyze the state
on paper using the various separability criteria. If no prior knowledge exists about the 
state, then it appears that there is no shortcut to such a quantum tomography 
experiment. In the more common situation in which we expect to have created
a certain state $\rho$, more efficient methods exist. 
A traditional method for detecting entanglement (employed for example in Ref. \cite{bouwmeester}) is to test for a violation of a Bell type inequality. We have indicated in this 
review that these tests are part of a larger framework of entanglement witnesses which exists both in the bipartite as well as in the multipartite setting. In Ref. \cite{lewpos2} the notion of an optimal entanglement witness was 
introduced; an entanglement witness $H$ is {\em optimal} if there exists no
other witness $H'$ which detects the same entanglement (the same states) 
as $H$ {\em and more}. These optimal witnesses are the ones that will be 
useful in detecting quantum entanglement. We will need the following 
additional definition:

\begin{defn}
A $\rho$-optimal witness $H_*$ for an entangled state $\rho$ is an optimal entanglement witness according to the definition in Ref. \cite{lewpos2} and among 
the optimal normalized witnesses, it is the best in detecting the entanglement of $\rho$, i.e. 
\beq
{\rm Tr}\, H_* \rho=\min_{\mbox{\small optimal }H} {\rm Tr}\, H \rho,  
\eeq
with $H$ normalized as
\beq
{\rm Tr} H=1.
\eeq
\label{defrhoopt}
\end{defn}

In the next Sections we will consider the optimal $\psi$-witness for pure states $\ket{\psi}$ and small systems, for mixed states in larger systems and 
for multipartite entangled states. In Section \ref{vinc} we consider how well these entanglement witnesses detect entanglement in the vicinity of the desired
entangled state.

\subsection{(Multipartite) pure states, small bipartite mixed states}
\label{decew}

Let us assume that we believe that the state of our multi- or bipartite 
quantum system is an entangled state $\ket{\psi}$ and let us assume that 
we are interested in detecting the entanglement of $\ket{\psi}$ over 
some bipartite A-B cut ${\mathcal H}_A \otimes {\mathcal H}_B$.
For pure states, there always exists a witness $H$ of the form 
\beq
H=a P+(1-a) ({\bf 1}\otimes T)(Q),
\label{normdec}
\eeq
where $P \geq 0$ and $Q \geq 0$, see Section \ref{ew}. 
An optimal witness (see Theorem 2, Ref. \cite{lewpos2}) in this class 
\footnote{We will only be optimizing amongst the decomposable witnesses 
to keep things as simple as possible.} has $a=0$ and has the property that the 
operator $Q$ has no product states in its range. To optimize with respect to the state $\ket{\psi}$ we choose $Q=\ket{\psi_{\mu_{min}}}\bra{\psi_{\mu_{min}}}$ where $\ket{\psi_{\mu_{min}}}$ is the eigenvector of $({\bf 1} \otimes T)(\ket{\psi}\bra{\psi})$ which has the smallest eigenvalue (which is negative). 
Such a witness is optimal in the sense of Ref. \cite{lewpos2}. The optimality of this choice with respect to $\ket{\psi}$ follows from 
\beq
{\rm Tr} \, ({\bf 1} \otimes T)(Q) \ket{\psi}\bra{\psi}={\rm Tr}\, Q\, ({\bf 1} \otimes T)(\ket{\psi}\bra{\psi}) \geq \lambda_{min},
\eeq
since $Q \geq 0$ and ${\rm Tr}\, Q =1$ due to normalization.

To see the explicit form of such a witness, we write $\ket{\psi}$
in the Schmidt decomposition $\ket{\psi}=\sum_i \sqrt{\lambda_i} \ket{a_i} \otimes \ket{b_i}$ and take the partial transpose in a fixed basis $\ket{0},\ket{1},\ldots$. We have 
\beq
({\bf 1} \otimes T)(\sum_{i,j} \sqrt{\lambda_i \lambda_j} \ket{a_i}\bra{a_j}
\otimes \ket{b_i}\bra{b_j})=\sum_{i,j} \sqrt{\lambda_i \lambda_j} \ket{a_i}\bra{a_j}
\otimes \ket{b_j^*}\bra{b_i^*},
\eeq
which has eigenvectors and corresponding eigenvalues $\{\ket{a_i,b_i^*},\lambda_i\}$ and for $i \neq j$, 
\beq
\left\{\frac{1}{\sqrt{2}}(\ket{a_i} \otimes \ket{b_j^*}\pm \ket{a_j} \otimes \ket{b_i^*}),\;\;\; \pm \sqrt{\lambda_i \lambda_j}\right\}.
\eeq
We choose the eigenvector with the most negative of the eigenvalues, let us 
call it $\mu_{min}=-\max_{i \neq j}\sqrt{\lambda_i \lambda_j}$. The optimal 
$\psi$-witness is equal to 
\beq
H_*=\frac{1}{2}(\ket{a_i,b_j}\bra{a_i,b_j}+\ket{a_j,b_i}\bra{a_j,b_i}-
\ket{a_i,b_i}\bra{a_j,b_j}-\ket{a_j,b_j}\bra{a_i,b_i}).
\eeq
for the pair $(i,j)$ corresponding to $\mu_{min}$. Let us take a simple 
example.

\begin{exmp}
For the state $\ket{\psi}=\cos \theta \ket{00}+\sin \theta \ket{11}$, the 
optimal witness is
\beq
H_*=\frac{1}{2}(\ket{0,1}\bra{0,1}+\ket{1,0}\bra{1,0}-\ket{0,0}\bra{1,1}-
\ket{1,1}\bra{0,0}),
\eeq
and $\mu_{min}=-\sqrt{\cos \theta \sin \theta}$. 
\label{simple_ex}
\end{exmp}

For bipartite quantum systems consisting of 2 qubits or 1 qutrit + 1 qubit, 
the entanglement witness is always decomposable, i.e. of the form 
$H=a P +(1-a)({\bf 1} \otimes T)(Q)$ (see Section \ref{ew}). As for pure states, the witness is optimal when $a=0$ and $Q$ has no product states in its range. To optimize for $\rho$ among such witnesses, we find the 
eigenvector $\ket{\psi_{\mu_{min}}}$ \footnote{This eigenvector is always 
entangled, since for all product states $\ket{\psi_A,\psi_B}$, $\bra{\psi_A,\psi_B}\, ({\bf 1}\otimes T)(\rho)\, \ket{\psi_A,\psi_B}={\rm Tr}\,\rho \,
({\bf 1}\otimes T)(\ket{\psi_A,\psi_B}\bra{\psi_A,\psi_B}) \geq 0$.} of $({\bf 1} \otimes T)(\rho)$ with the smallest eigenvalue $\mu_{min}$ and we choose 
\beq
H_*=({\bf 1} \otimes T)(\ket{\psi_{\mu_{min}}}\bra{\psi_{\mu_{min}}}).
\eeq

\subsection{Mixed State Entanglement in Higher Dimensions}
Let $\rho$ be a (multipartite) mixed state in dimensions more than ${\mathcal H}_2 \otimes {\mathcal H}_2$ or ${\mathcal H}_2 \otimes {\mathcal H}_3$ whose
entanglement we wish to detect over a bipartite cut A-B. If $\rho$ violates the Peres-Horodecki criterion then the methods in the previous Section can be applied to determine an optimal entanglement witness. When 
$\rho$ has the PPT property and is believed to be entangled, then $\rho$ has 
bound entanglement and we will need an indecomposable entanglement witness
for $\rho$ (see Section \ref{ew}). In Ref. \cite{lewpos2} a method was 
developed to optimize a given indecomposable entanglement witness. The problem 
is a lot harder than for decomposable witnesses: a generic form for 
an optimal witness is not known. Nonetheless we can sketch a procedure for finding an optimal indecomposable witness which detects $\rho$:
\begin{enumerate}
\item Find a decomposable witness $H$, Eq. (\ref{genform22}), for which 
${\rm Tr}\, H \rho=0$ with ${\rm Tr}\, H=1$.
\item Choose as a starting point the indecomposable witness
$H'=(H-\eps {\bf 1})/(1-\eps d)$ where $\eps=\inf_{\psi_1,\psi_2} \bra{\psi_1,\psi_2}\, H\, \ket{\psi_1,\psi_2}$ which ensures that $H'$ is a (normalized) witness. Here 
$d$ is the total dimension of the quantum system.
\item Optimize $H'$ to $H_*$ with the methods in Ref. \cite{lewpos2}, taking 
into account the optimality with respect to $\rho$, see Definition \ref{defrhoopt}. The optimized witness relates to $H'$ as  
\beq
H_*=(H'-\lambda_* D_*)/(1-\lambda_*),
\eeq
where $\lambda_*$ is a constant depending on $D_*$ and $H'$ (see Eq. (13) in 
Ref. \cite{lewpos2}) and where $D_*$ is a normalized decomposable witness, Eq. (\ref{normdec}), with the property that for all product states $\ket{\psi_1,\psi_2}$ such that (1) \linebreak
$\bra{\psi_1,\psi_2}\, H' \,\ket{\psi_1,\psi_2}=0$ we have 
$\bra{\psi_1, \psi_2}\, D_* \, \ket{\psi_1,\psi_2}=0$, (2) $\lambda_*$ and 
$D_*$ are such that $H_*$ is an optimal indecomposable entanglement 
witness (no more decomposable witnesses can be subtracted from it) and
(3) when under constraints (1) and (2) there is still freedom of choice in 
$\lambda_*$ and $D_*$, we choose an $H_*$ which is optimal with respect to 
$\rho$, i.e. ${\rm Tr}\, H_* \rho$ is minimal.
\end{enumerate}

It is not guaranteed that this procedure will lead to a $\rho$-optimal 
indecomposable witness according to Definition \ref{defrhoopt}. The method
does however ensure that the witness is optimal as well as detecting 
the entanglement of $\rho$.

\subsection{Multipartite (Bound) Entanglement}

The entanglement witness framework carries over to multipartite states.
This framework is particularly useful when entanglement is to be detected
in a bound entangled state such as the first example $\rho_{\bf Shifts}$, Eq. (\ref{shiftbe}), in Section \ref{bemulti} which is separable over all bipartitions. For this state an entanglement witness was found in 
Ref. \cite{nsep:horodecki}, in analogy to the construction in Ref. \cite{terhalposmap}. Even though various Bell inequalities and entanglement witnesses 
exist for multipartite entanglement (Sections \ref{bellmulti} and \ref{ewmulti}), a more elaborate theory as in the bipartite case is still lacking. We would also like to refer the reader to Ref. \cite{dur_detect} for alternative ideas on the experimental detection of entanglement in multiqubit states.

\subsection{The vicinity of the trial state}
\label{vinc}

Often the quantum system under consideration will not exactly be 
in the desired state $\rho$. It is therefore essential that the entanglement 
witness that we chose is robust, in the sense that it 
detects as much entanglement as possible in the neighborhood of $\rho$ (provided that the neighborhood is entangled). This property is guaranteed in the following 
manner. Let $\rho'$ be all states in the 'vicinity' of $\rho$, namely for a given $\eps$, $\Delta=\rho'-\rho$
\beq
||\Delta||_1 \leq \epsilon,
\eeq
where $||.||_1$ is the tracenorm, i.e. $||A ||_1={\rm Tr} \sqrt{A^{\dagger} A}$. Letting $H_*$ be the optimal $\rho$-witness, we have
\beq
{\rm Tr}\, H_* \rho'={\rm Tr}\, H_* \rho +
{\rm Tr}\, H_* \Delta.
\eeq
The last quantity can be bounded, using the Schwarz inequality
$|{\rm Tr}\, A^{\dagger} B| \leq \sqrt{{\rm Tr}\,A^{\dagger} A} \sqrt{{\rm Tr}\,B^{\dagger} B}$ and also $\sqrt{{\rm Tr}\,A^{\dagger} A} \leq {\rm Tr}\,\sqrt{A^{\dagger} A}$. This gives 
\beq
{\rm Tr}\, H_* \rho- \eps\, \sqrt{{\rm Tr}H_*^2} \leq {\rm Tr}\, H_* \rho' \leq{\rm Tr}\, H_* \rho+\eps\, \sqrt{{\rm Tr}H_*^2}.
\eeq
For the optimal decomposable entanglement witnesses given in Section \ref{decew} we have $\sqrt{{\rm Tr}H_*^2}=1$, irrespective of the dimension of the system. Thus 
when $\rho'$ is close to the state $\rho$, the optimal witness $H_*$ for $\rho$ will also detect the entanglement in $\rho'$.
We will now consider a specific example in which all states in a certain class are detected by one entanglement witness:

\begin{exmp} Let $\ket{\Psi}$ be any maximally entangled pure state in a bipartite space of total dimension $d$. Instead of $\ket{\Psi}$ the physical state of our quantum system is
\beq
\rho_p=p \ket{\Psi}\bra{\Psi}+\frac{1-p}{d}{\bf 1}. 
\eeq
As long as $({\bf 1}\otimes T)(\rho_p) \not \geq 0$, the optimal $\Psi$-witness
$H_*=({\bf 1}\otimes T)(\ket{\psi_{\mu_{min}}}\bra{\psi_{\mu_{min}}})$ will detect the entanglement of $\rho_p$. The eigenvector with the 
minimal eigenvalue of $({\bf 1}\otimes T)(\rho_p)$ is the same for all $p \in [0,1]$.
Furthermore the density matrix $\rho_p$ has the property that it is separable 
{\em if and only if} it satisfies the Peres-Horodecki criterion \cite{filterhor}. Therefore the entanglement in all (entangled) $\rho_p$ is witnessed by $H_*$, since ${\rm Tr} \,H_* \rho_p$ is negative as long as $\rho_p$ violates the Peres-Horodecki criterion.
\end{exmp}

\subsection{Measurement of the witness}

In principle, the entanglement witness method has the advantage that only 
one observable, the entanglement witness, needs to be measured. In practice, the measurement of this observable may be done by a series of local measurements. 
Consider for example the entanglement witness of Example \ref{simple_ex}, 
in terms of the Pauli-matrices it reads
\beq
H_*=\frac{1}{4}(I \otimes I+Y \otimes Y-X \otimes X-Z \otimes Z). 
\label{wit}
\eeq
A measurement of all the Pauli matrices on both sides is needed to 
measure this observable by local measurements. At this point the advantage 
over basic state tomography becomes somewhat questionable. The entanglement witness will in general be a nonlocal observable. But if we allow for nonlocal
measurements, or rotate the state by a nonlocal rotation $U$ prior to 
testing such that $U H_* U^{\dagger}$ is local, there exists a real advantage over state tomography. A CNOT gate where the first qubit is the control and the second is the target, will rotate $H_*$ in Eq. (\ref{wit}) to a product of operators 
\beq
{\rm CNOT}_{A \rightarrow B}\, H_* \,{\rm CNOT}_{A \rightarrow B}^{\dagger}=\frac{1}{4}({\bf 1}-X)\otimes ({\bf 1}-Z)).
\eeq
Thus some quantum computation power is needed to measure the entanglement witness efficiently. For large systems one may ask whether such quantum computation can be performed efficiently, in polynomial time in the number of qubits, for states which can be built efficiently. With this open question we will conclude our review.

\section{Acknowledgments}Thanks to Patrick Hayden for suggesting 
an improved counterexample in Section \ref{sepcrit}, to Micha{\l} 
and Pawe{\l} Horodecki for discussions on the entropic inequalities, to Daniel Gottesman for discussions on homomorphisms and positive maps and to David 
DiVincenzo for reading through this manuscript.

\appendix

\section{Indecomposable Positive Linear Maps and Hilbert's 17th problem}
\label{hilbert17}

In Ref. \cite{choi:biquad} it was shown how certain indecomposable positive 
linear maps present answers (in the negative) to Hilbert's 17th problem. Hilbert's
17th problem asks whether all positive semidefinite homogeneous polynomials
are sums of squares of homogeneous polynomials (see Ref. \cite{reznick} for 
some history on the problem). Indecomposable positive maps with real coefficients present counterexamples for polynomials which are real biquadratic forms. The construction is the following.

Given is a positive indecomposable map ${\mathcal L}\colon 
B({\mathcal H}_n) \rightarrow B({\mathcal H}_m)$ which is known not to be 
completely positive. Furthermore, the coefficients of ${\mathcal L}$ in some 
fixed basis $\{\ket{i}\}$ are real and therefore ${\mathcal L}$ maps real (symmetric)
matrices in $B({\mathcal H}_n)$ onto real symmetric matrices in $B({\mathcal H}_m)$. Note that matrix transposition acts as the identity on real symmetric matrices and therefore we restrict ourselves to indecomposable maps.

We define a positive semidefinite symmetric biquadratic form in the following way:
\beq
F(x_1,\ldots, x_n; y_1, \ldots y_m)=\bra{y}\,({\mathcal L}(\ket{x}\bra{x})\, \ket{y}, 
\eeq
where $\ket{x}$ and $\ket{y}$ are unnormalized vectors with  coefficients $\ket{x}=\sum_i x_i \ket{i}$ and $\ket{y}=\sum_i y_i \ket{i}$, $x_i \in {\bf R}$ and $y_i \in {\bf R}$.
The positivity of the map ${\mathcal L}$ guarantees the positive semidefiniteness of $F(x_1,\ldots, x_n; y_1, \ldots y_m)$ for all $x_1,\ldots,x_n, y_1,\ldots, y_m$, or 
\beq
F(x_1,\ldots, x_n; y_1, \ldots y_m)=\sum_{i,j,k,l} {\mathcal L}_{ij,kl} x_i x_j
y_k y_l \geq 0.
\eeq
Now we pose Hilbert's question: is every positive semidefinite symmetric biquadratic form a sum of squares, i.e. 
\beq
F(x_1,\ldots, x_n; y_1, \ldots y_m) \stackrel{?}{=} \sum_t 
(G_t(x_1,\ldots, x_n; y_1, \ldots y_m))^2, 
\label{sumofsquares}
\eeq
where $G_t(x_1,\ldots, x_n; y_1, \ldots y_m)$ is a symmetric bilinear form, i.e.
\beq 
G_t(x_1,\ldots, x_n; y_1, \ldots y_m)=\sum_{i,j} g_{ij}^t x_i y_j.
\eeq

Assume that the equality in Eq. (\ref{sumofsquares}) holds. We define a set of real operation elements $A_t$ with
\beq
\bra{j}\, A_t\, \ket{i}=g_{ij}^t, 
\eeq
so that 
\beq
\sum_t (G_t(x_1,\ldots, x_n; y_1, \ldots y_m))^2=\sum_t \bra{y}\, A_t \ket{x}\bra{x} A_t^{T} \ket{y}.  
\eeq
This would imply that the map ${\mathcal L}$ has a decomposition in terms 
of the operation elements $A_t$ and therefore ${\mathcal L}$ is a completely positive map. Conversely, when a map ${\mathcal L}$ is completely positive, the 
corresponding positive semidefinite symmetric biquadratic form is a sum of 
squares.
Instead of the positive indecomposable map ${\mathcal L}$, we could have defined the biquadratic form in terms of an (indecomposable) entanglement witness $H$ with real coefficients:
\beq
F(x_1,\ldots,x_n;y_1,\ldots,y_m)=\bra{y,x}\,H\,\ket{y,x} \geq 0,
\eeq
for all real vectors $\ket{y,x}$. It was shown how to construct (real-valued) entanglement witnesses for every (real) unextendible product bases (UPB) in Ref. \cite{terhalposmap}; furthermore the graph-theoretic 
UPB construction in Ref. \cite{lovalon} always has a realization with real vectors.

\bibliographystyle{halpha}
\bibliography{refs}

\newcommand{\etalchar}[1]{$^{#1}$}
\begin{thebibliography}{BDM{\etalchar{+}}99}

\bibitem[AL]{lovalon}
N.~Alon and L.~Lov\'asz.
\newblock Unextendible product bases.
\newblock Manuscript Jan. 2000, to appear in Journal of Combinatorial Theory,
  Ser. A.

\bibitem[And94]{majando}
T.~Ando.
\newblock Majorizations and inequalities in matrix theory.
\newblock {\em Linear Algebra and Its Applications}, 199:17--67, 1994.

\bibitem[BBC{\etalchar{+}}93]{tele}
C.H. Bennett, G.~Brassard, C.~Cr{\'e}peau, R.~Jozsa, A.~Peres, and W.K.
  Wootters.
\newblock Teleporting an unknown quantum state via dual classical and
  {Einstein-Podolsky-Rosen} channels.
\newblock {\em Phys. Rev. Lett.}, 70:1895--1899, 1993.

\bibitem[BBPS96]{bbps}
C.H. Bennett, H.J. Bernstein, S.~Popescu, and B.~Schumacher.
\newblock Concentrating partial entanglement by local operations.
\newblock {\em Phys. Rev. A}, 53:2046--2052, 1996.

\bibitem[BDM{\etalchar{+}}99]{upb1}
C.H. Bennett, D.P. DiVincenzo, T.~Mor, P.W. Shor, J.A. Smolin, and B.M. Terhal.
\newblock Unextendible product bases and bound entanglement.
\newblock {\em Phys. Rev. Lett.}, 82:5385--5388, 1999, quant-ph/9808030.

\bibitem[BDSW96]{bdsw}
C.H. Bennett, D.P. DiVincenzo, J.A. Smolin, and W.K. Wootters.
\newblock Mixed state entanglement and quantum error correction.
\newblock {\em Phys. Rev. A}, 54:3824--3851, 1996.

\bibitem[Bel64]{bell}
J.S. Bell.
\newblock On the {Einstein-Podolsky-Rosen} paradox.
\newblock {\em Physics}, 1:195--200, 1964.

\bibitem[BPD{\etalchar{+}}99]{bouwmeester}
D.~Bouwmeester, J.-W. Pan, M.~Daniell, H.~Weinfurter, and A.~Zeilinger.
\newblock Observation of three-photon {G}reenberger-{H}orne-{Z}eilinger
  entanglement.
\newblock {\em Phys. Rev. Lett.}, 82:1345--1349, 1999, quant-ph/9810035.

\bibitem[BPR{\etalchar{+}}00]{multibenn}
C.~H. Bennett, S.~Popescu, D.~Rohrlich, J.~A. Smolin, and A.~V. Thapliyal.
\newblock Exact and asymptotic measures of multipartite pure state
  entanglement.
\newblock {\em Phys. Rev. A}, 63:012307, 2000, quant-ph/9908073.

\bibitem[CAG99]{cag}
N.~J. Cerf, C.~Adami, and R.~M. Gingrich.
\newblock Quantum conditional operator and a criterion for separability.
\newblock {\em Phys. Rev. A}, 60:893--898, 1999, quant-ph/9710001.

\bibitem[Cho75]{choi:biquad}
M.-D. Choi.
\newblock Positive semidefinite biquadratic forms.
\newblock {\em Linear Algebra and Its Applications}, 12:95--100, 1975.

\bibitem[CHSH69]{chsh}
J.F. Clauser, M.A. Horne, A.~Shimony, and R.A. Holt.
\newblock Proposed experiment to test local hidden-variable theories.
\newblock {\em Phys. Rev. Lett.}, 23:880, 1969.

\bibitem[DC]{dur_detect}
W.~D{\"ur} and J.I. Cirac.
\newblock Multiparticle entanglement and its experimental detection.
\newblock quant-ph/0011025.

\bibitem[DC00]{durcirac}
W.~D{\"u}r and J.I. Cirac.
\newblock Classification of multi-qubit mixed states: separability and
  distillability properties.
\newblock {\em Phys. Rev. A}, 61:042314, 2000, quant-ph/9911044.

\bibitem[DCLB00]{nptnond2}
W.~D{\"u}r, J.I. Cirac, M.~Lewenstein, and D.~Bruss.
\newblock Distillability and partial transposition in bipartite systems.
\newblock {\em Phys. Rev. A}, 61:062313, 2000, quant-ph/9910022.

\bibitem[DMS{\etalchar{+}}]{upb2}
D.P. DiVincenzo, T.~Mor, P.W. Shor, J.A. Smolin, and B.M. Terhal.
\newblock Unextendible product bases, uncompletable product bases and bound
  entanglement.
\newblock accepted by Comm. Math. Phys., quant-ph/9908070.

\bibitem[DP97]{divperes}
D.P. DiVincenzo and A.~Peres.
\newblock Quantum codewords contradict local realism.
\newblock {\em Phys. Rev. A}, 55:4089--4092, 1997, quant-ph/9611011.

\bibitem[DSS{\etalchar{+}}00]{nptnond1}
D.P. DiVincenzo, P.W. Shor, J.A. Smolin, B.M. Terhal, and A.V. Thapliyal.
\newblock Evidence for bound entangled states with negative partial transpose.
\newblock {\em Phys. Rev. A}, 61:062312/1--13, 2000, quant-ph/9910026.

\bibitem[DVC00]{durvidalcirac}
W.~D{\" u}r, G.~Vidal, and J.I. Cirac.
\newblock Three qubits can be entangled in two inequivalent ways.
\newblock {\em Phys. Rev. A}, 62:062314, 2000, quant-ph/0005115.

\bibitem[GBP98]{bech:bell}
N.~Gisin and H.~Bechmann-Pasquinucci.
\newblock Bell inequality, {B}ell states and maximally entangled states for n
  qubits.
\newblock {\em Physics Letters A}, 246:1--6, 1998.

\bibitem[GC99]{gott&chuang}
D.~Gottesman and I.~Chuang.
\newblock Demonstrating the viability of universal quantum computation using
  teleportation and single qubit operations.
\newblock {\em Nature}, 402(6760):390--393, 1999, quant-ph/9908010.

\bibitem[GPV]{galvao}
E.F. Galvao, M.B Plenio, and S.~Virmani.
\newblock Tripartite entanglement and quantum relative entropy.
\newblock quant-ph/0008089.

\bibitem[HH]{privhor}
M.~Horodecki and P.~Horodecki.
\newblock Private Communication.

\bibitem[HH99]{filterhor}
M.~Horodecki and P.~Horodecki.
\newblock Reduction criterion of separability and limits for a class of
  distillation protocols.
\newblock {\em Phys. Rev. A}, 59:4206--4216, 1999, quant-ph/9708015.

\bibitem[HHH]{nsep:horodecki}
M.~Horodecki, P.~Horodecki, and R.~Horodecki.
\newblock Separability of n-particle mixed states: necessary and sufficient
  conditions in terms of linear maps.
\newblock quant-ph/0006071.

\bibitem[HHH96]{nec_horo}
M.~Horodecki, P.~Horodecki, and R.~Horodecki.
\newblock Separability of mixed states: necessary and sufficient conditions.
\newblock {\em Physics Letters A}, 223:1--8, 1996, quant-ph/9605038.

\bibitem[HHH97]{dist2d}
M.~Horodecki, P.~Horodecki, and R.~Horodecki.
\newblock Inseparable two spin 1/2 density matrices can be distilled to a
  singlet form.
\newblock {\em Phys. Rev. Lett.}, 78:574--577, 1997, quant-ph/9607009.

\bibitem[HHH98]{pptnodist}
M.~Horodecki, P.~Horodecki, and R.~Horodecki.
\newblock Mixed state entanglement and distillation: is there a `bound'
  entanglement in nature?
\newblock {\em Phys. Rev. Lett.}, 80:5239--5242, 1998, quant-ph/9801069.

\bibitem[HHH00]{review:horodecki}
M.~Horodecki, P.~Horodecki, and R.~Horodecki.
\newblock {\em Mixed-state entanglement and quantum communication in {\em
  Quantum Information --Basic Concepts and Experiments. Ed. G. Alber and M.
  Weiner}}.
\newblock Springer, Berlin, 2000.

\bibitem[HLVC00]{berank}
P.~Horodecki, M.~Lewenstein, G.~Vidal, and I.~Cirac.
\newblock Operational criterion and constructive checks for the separability of
  low rank density matrices.
\newblock {\em Phys. Rev. A}, 62:032310, 2000, quant-ph/0002089.

\bibitem[Hor97]{bepawel}
P.~Horodecki.
\newblock Separability criterion and inseparable mixed states with positive
  partial transposition.
\newblock {\em Physics Letters A}, 232:333--339, 1997, quant-ph/9703004.

\bibitem[HSTT]{norank2}
P.~Horodecki, J.A. Smolin, B.M. Terhal, and A.V. Thapliyal.
\newblock Rank two bound entangled states do not exist.
\newblock quant-ph/9910122.

\bibitem[Jam72]{jamiolkowski}
A.~Jamio\l{}kowski.
\newblock Linear transformations which preserve trace and positive
  semidefiniteness of operators.
\newblock {\em Rev. of Mod. Phys.}, 3:275--278, 1972.

\bibitem[JB00]{janzing:ew}
D.~Janzing and Th. Beth.
\newblock Fragility of a class of highly entangled states with n qubits.
\newblock {\em Phys. Rev. A}, 61:052308, 2000, quant-ph/9907042.

\bibitem[KLM]{klm}
E.~Knill, R.~Laflamme, and G.~Milburn.
\newblock Efficient linear optics quantum computation.
\newblock quant-ph/0006088.

\bibitem[KLMT00]{nmrcat}
E.~Knill, R.~Laflamme, R.~Martinez, and C.-H. Tseng.
\newblock An algorithmic benchmark for quantum information processing.
\newblock {\em Nature}, 404:368--370, 2000, quant-ph/9908051.

\bibitem[LKCH]{lewpos2}
M.~Lewenstein, B.~Kraus, J.I. Cirac, and P.~Horodecki.
\newblock Optimization of entanglement witnesses.
\newblock quant-ph/0005014.

\bibitem[LKHC]{lewpos1}
M.~Lewenstein, B.~Kraus, P.~Horodecki, and J.I. Cirac.
\newblock Characterization of separable states and entanglement witnesses.
\newblock quant-ph/0005112.

\bibitem[LPSW]{linden:rev}
N.~Linden, S.~Popescu, B.~Schumacher, and M.~Westmoreland.
\newblock Reversibility of local transformations of multiparticle entanglement.
\newblock quant-ph/9912039.

\bibitem[Mer90a]{mermin_multi}
N.D. Mermin.
\newblock Extreme quantum entanglement in a superposition of macroscopically
  distinct states.
\newblock {\em Phys. Rev. Lett.}, 65:1838--1840, 1990.

\bibitem[Mer90b]{mermin}
N.D. Mermin.
\newblock Quantum mysteries revisited.
\newblock {\em Am. J. Phys.}, 58:731--734, 1990.

\bibitem[NK]{niel_kemp}
M.A. Nielsen and J.~Kempe.
\newblock Separable states are more disordered globally than locally.
\newblock quant-ph/0011117.

\bibitem[Per93]{peresbook}
A.~Peres.
\newblock {\em Quantum Theory: Concepts and Methods}.
\newblock Kluwer Academic Publishers, 1993.

\bibitem[Per96]{Asher96}
A.~Peres.
\newblock Separability criterion for density matrices.
\newblock {\em Phys. Rev. Lett.}, 77:1413--1415, 1996.

\bibitem[Rai99]{rainsdist}
E.M. Rains.
\newblock Rigorous treatment of distillable entanglement.
\newblock {\em Phys. Rev. A}, 60:173--178, 1999.

\bibitem[RB]{rausbrieg}
R.~Rausschendorf and H.~Briegel.
\newblock Quantum computing via measurements only.
\newblock quant-ph/0010033.

\bibitem[Rez]{reznick}
B.~Reznick.
\newblock Some concrete aspects of {H}ilbert's 17th problem.
\newblock http://www.math.uiuc.edu/Reports/reznick/98-002.html.

\bibitem[SCK{\etalchar{+}}]{fermions_ent}
J.~Schliemann, J.I. Cirac, M.~Ku{\'s}, M.~Lewenstein, and D.~Loss.
\newblock Quantum correlations in two-fermion systems.
\newblock quant-ph/0012094.

\bibitem[Smo]{smolin}
J.A. Smolin.
\newblock A four-party unlockable bound-entangled state, quant-ph/0001001.

\bibitem[SSTa]{sst:bip}
P.W. Shor, J.A. Smolin, and B.M. Terhal.
\newblock Evidence for nonadditivity of bipartite distillable entanglement.
\newblock quant-ph/0010054.

\bibitem[SSTb]{sst:act}
P.W. Shor, J.A. Smolin, and A.V. Thapliyal.
\newblock Superactivation of bound entanglement, quant-ph/0005117.

\bibitem[TD00]{terhal&divincenzo:equilib}
B.~M. Terhal and D.~DiVincenzo.
\newblock Problem of equilibration and the computation of correlation functions
  on a quantum computer.
\newblock {\em Phys. Rev. A}, 61:22301, 2000, quant-ph/9810063.

\bibitem[Ter]{terhalposmap}
B.M. Terhal.
\newblock A family of indecomposable positive linear maps based on entangled
  quantum states.
\newblock Accepted by Lin. Alg and Its Appl., quant-ph/9810091.

\bibitem[Ter00]{terhalbell}
B.M. Terhal.
\newblock Bell inequalities and the separability criterion.
\newblock {\em Physics Letters A}, 271:319--326, 2000, quant-ph/9911057.

\bibitem[Weh78]{wehrl}
A.~Wehrl.
\newblock General properties of entropy.
\newblock {\em Rev. of Mod. Phys.}, 50:221--260, 1978.

\bibitem[Wer89]{werner:lhv}
R.F. Werner.
\newblock Quantum states with {Einstein-Podolsky-Rosen} correlations admitting
  a hidden-variable model.
\newblock {\em Phys. Rev. A}, 40:4277--4281, 1989.

\bibitem[Wor76]{woronowicz}
S.~L. Woronowicz.
\newblock Positive maps of low dimensional matrix algebras.
\newblock {\em Rep. in Math. Phys.}, 10:165--183, 1976.

\bibitem[WW00]{werner:wolf}
R.F. Werner and M.M. Wolf.
\newblock Bell's inequalities for states with positive partial transpose.
\newblock {\em Phys. Rev. A}, 61:062102, 2000.

\end{thebibliography}

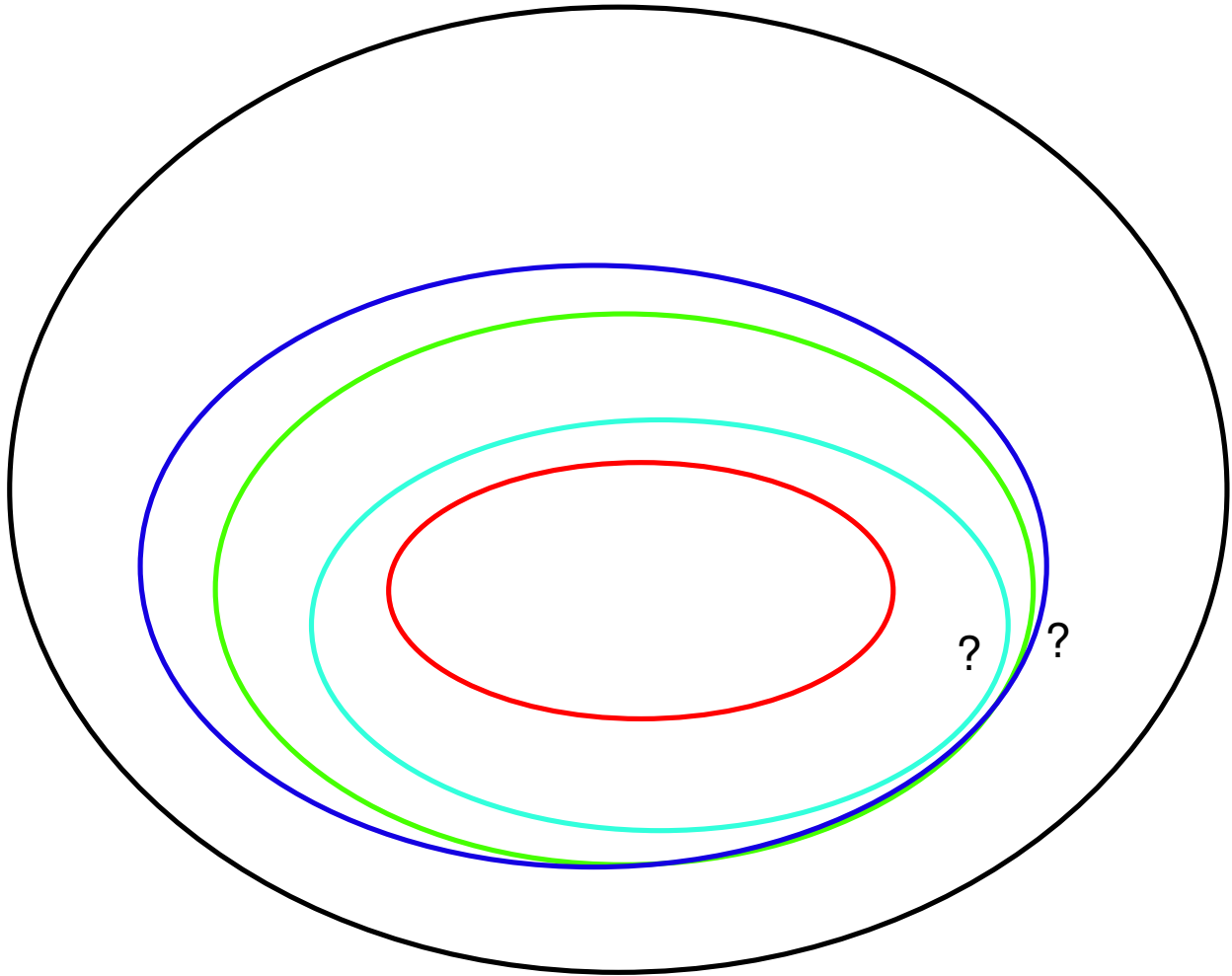
\end{document}